\documentclass[aps,prd,twocolumn,showpacs,preprintnumbers,amsmath,amssymb]{revtex4}

 \def\frac#1#2{{#1\over #2}}

\def\be{\begin{equation}}
\def\ee{\end{equation}}
\def\ba{\begin{eqnarray}}
\def\ea{\end{eqnarray}}

 \def\we{\wedge}
 
 \def\f {\frac}
 \def\ti{\tilde}
 \def\ap{\alpha}
  \def\al{\alpha'}

 \def\no{\nonumber \\}

 \def\lb{\rangle}

 \def\ov{\overline}

\usepackage{color}
\usepackage{graphicx}
\usepackage{dcolumn}
\usepackage{bm}
\usepackage[hypertex]{hyperref}

\begin{document}

\preprint{IPMU10-0126}

\title{Topological Insulators and Superconductors from String Theory}

\author{Shinsei Ryu}
\affiliation{
Department of Physics, University of California, Berkeley, CA 94720, USA
            }

\author{Tadashi Takayanagi}
\affiliation{
Institute for the Physics and Mathematics of the Universe (IPMU),
University of Tokyo, Kashiwa, Chiba 277-8582, Japan
            }

\date{\today}

\begin{abstract}
Topological insulators and superconductors
in different spatial dimensions and with different discrete symmetries have been fully classified
recently,
revealing a periodic structure
for the pattern of possible types of
topological insulators and supercondutors,
both in terms of spatial dimensions
and in terms of symmetry classes.
It was proposed that K-theory is behind the periodicity.
On the other hand,
D-branes,
a solitonic object in string theory,
are also known to be classified by K-theory.
In this paper, by inspecting low-energy effective field theories
realized by two parallel D-branes,
we establish a one-to-one correspondence between
the K-theory classification of
topological insulators/superconductors and D-brane charges.
In addition,
the string theory realization of topological insulators and superconductors comes naturally with
gauge interactions, and
the Wess-Zumino term of the D-branes gives rise to a gauge field theory
of topological nature,
such as ones with the Chern-Simons term or the $\theta$-term
in various dimensions.
This sheds light on
topological insulators and superconductors beyond non-interacting systems,
and the underlying topological field theory
description thereof.
In particular, our string theory realization includes
the honeycomb lattice Kitaev model in two spatial dimensions,
and its higher-dimensional extensions.
Increasing the number of D-branes naturally leads to
a realization of topological insulators and superconductors
in terms of holography (AdS/CFT).
\end{abstract}

\maketitle


\section{Introduction}

The integer quantum Hall effect (IQHE) is one of the most striking phenomenon
in the $d=2$-dimensional electron system under a strong magnetic field,
and has been one of the central topics in condensed matter physics.
As well-known, the Hall conductance is quantized when
the electronic ground state has a non-trivial topological structure.

With the recent discovery of the quantum spin Hall effect (QSHE) in $d=2$
and the $\mathbb{Z}_2$ topological insulator in $d=3$
\cite{KaneMele,Roy,moore07,bernevig06,konig07,Fu06_3Da, Fu06_3Db, Qi2008,
hasan07,Hsieh09, Xia09, Hsieh09b, Chen09, review},
it has become clear that
topological phases can exist in a much wider context,
i.e., in the spatial dimension other then $d=2$,
and without strong time-reversal symmetry (TRS) breaking
by a magnetic field.
The QSHE and $d=3$-dimensional $\mathbb{Z}_2$ topological insulator
can be thought of as a close cousin of the IQHE,
but different from the IQHE in many essential ways:
these states can exist only when time-reversal symmetry is respected,
and can be either two- or three-dimensional.
Furthermore, $\mathbb{Z}_2$ topological insulators are characterized by
a binary topological number,
unlike the integral Hall conductivity in the IQHE.
Recent experiments confirmed HgTe
quantum well and Bismuth-related materials,
both of which have strong spin-orbit coupling,
indeed realize
such non-trivial $\mathbb{Z}_2$ topological phases.

The Bloch wavefunctions in the IQHE or
the $\mathbb{Z}_2$ topological insulator in $d=2$ (the QSHE) and in $d=3$
have non-trivial topology, detected by an integer or a binary topological number.
For superconductors (superfluid),
at least within the BCS meanfield theory,
it is possible the wavefunctions of fermionic Bogoliubov quasiparticles
carry non-trivial topological characters,
very much the same way as the electronic wavefunctions in
the IQHE or the $\mathbb{Z}_2$ topological insulator,
in particular when fully gapped
(i.e., a quasiparticle gap opens for entire momentum space);
such superconductors (superfluid) can be called topological superconductor (superfluid).
A well-known example of such topological superconductor is
the $d=2$-dimensional chiral $p$-wave superconductor
which has a $p_x+ {i}p_y$-wave superconducting order parameter.

For non-interacting fermions,
an exhaustive classification of topological insulators (TIs) and superconductors (TSCs)
is proposed in Refs.\ \cite{SRFL,Kitaev}:
TIs/TSCs are classified in terms of spatial dimensions $d$
and the $10=2+8$ symmetry classes (two ``complex'' and eight ``real'' classes),
each characterized by presence/absence of discrete symmetries
such as time-reversal symmetry (TRS or T),
particle-hole symmetry (PHS or C), and
chiral (or sublattice) symmetry (SLS or S).
(Table \ref{charge}).
In relativistic field theories,
the first two are the same as the usual T and C symmetries.
For any system which has both T and C symmetries,
S symmetry (SLS) is realized as a product C$\cdot$T,
while it can exist on its own, even without T and C symmetries.
The ten symmetry classes are in one-to-one correspondence
to the Riemannian symmetric spaces without exceptional series
(described in Table \ref{cartan})
{\cite{Zirnbauer96, Altland97, Huckleberry2005, Helgason1978},
and to K-theory classifying spaces \cite{Kitaev,HoravaFermiSurfKtheory}.
For example, the integer QHE, QSHE,
and $\mathbb{Z}_2$ TI
are a topologically non-trivial state
belonging to class A ($d=2$), AII ($d=2$), and AII ($d=3$), respectively.

The classification reveals a periodicity both in spatial dimensions $d$
and in symmetry classes, and hence is often called
``periodic table'' of TIs/TSCs.
Not only it incorporate many of previously known topological phases,
it also predicted new topological phases;
e.g.,
the B phase of $^3$He was newly identified as a topological phase;
the existence of the $d=3$-dimensional topological singlet superconductor
was predicted, and verified by an explicit construction of
a lattice model of the BCS superconductor
\cite{SchnyderRyuLudwig2009}.

The complete classification of non-interacting TIs and TSCs opens up a number of further questions,
most interesting among which are interaction effects\cite{interactions}:
Do non-interacting topological phases continue to exist
in the presence of interactions?
Can interactions give rise to novel topological phases other than non-interacting TIs/TSCs?
What is a topological field theory underlying TIs/TSCs, which can potentially
describe TIs/TSCs beyond non-interacting examples?, etc.

On the other hand, the ten-fold classification of TIs/TSCs reminds us of D-branes,
which are fundamental objects in string theory,
and are also classified by K-theory \cite{WittenK,Hor}
(Table \ref{dbrane}) via the open string tachyon condensation \cite{SenSO,Sen}.
It is then natural to speculate a possible connection between TIs/TSCs and
of D-branes
\cite{Karch}.
In this paper, we propose a systematic construction of TIs/TSCs
in terms of various systems composed of two kinds of D-branes (D$p$- and D$q$-branes),
possibly with an orientifold plane (O-plane).
Besides the appealing mathematical similarity between TIs/TSCs and D-branes,
realizing TIs/TSCs in string theory has a number of merits, since
string theory and D-branes are believed
to be rich enough to reproduce many types of field theories and interactions
in a fully consistent and UV complete way.
Indeed, our string theory realizations of TIs/TSCs give rise to
massive fermion spectra,
which are in one-to-one correspondence with
the ten-fold classification of TIs/TSCs,
and come quite naturally
with gauge interactions.
These systems, while interacting,
are all topologically stable, as protected by the K-theory charge of D-branes.
We thus make a first step toward understanding interacting TIs/TSCs.
Our approach also reveals the connection between
the number of symmetry classes of TIs/TSCs
and
the critical dimension of superstring theory (=10),
via the Bott periodicity of K-theory.

In D$p$-D$q$-systems,
massive fermions arise as an open string excitation between the two D-branes.
The distance between the branes corresponds to the mass of fermions.
Open strings ending on the same D-branes give rise to a gauge field,
which we call $A_\mu$ (D$p$) and $\tilde{A}_\mu$ (D$q$)
with gauge group $G$ and $\tilde{G}$, respectively,
and couple to the fermions (they are identified as in Table \ref{gauge}).
These two gauge fields play different roles in our construction:
The gauge field $A_\mu$ ``measures'' the K-theory charge of the D$q$-brane,
and in that sense it can be interpreted as an ``external'' gauge field.
In this picture, the D$q$-brane charge is identified with the
topological (K-theory) charge of TIs/TSCs.
On the other hand,
$\tilde{A}_\mu$ is an internal degree of freedom on the D$q$-brane.

The massive fermions can be integrated out, yielding
the description of the topological phase in terms of the gauge fields.
The resulting effective field theory comes with terms of topological nature,
such as the Chern-Simons (CS) or the $\theta$-terms.
In our string theory setup,
they can be read off from the Wess-Zumino (WZ) action of the D-branes,
by taking one of the D-branes as a background for the other.
In our construction of the QSHE (i.e. AII d=2),
the brane system consists of
a D$p$-brane and a D$q$-$\overline{\mathrm{D}}q$ system,
where $\overline{\mathrm{D}}q$ denotes an anti D$q$-brane,
which has the opposite Ramond-Ramond charge
to the D$q$-brane.
The D$p$-brane has an $\mathrm{SU}(2)$ gauge field
and the D$q$-$\overline{\mbox{D}}q$ system has a $\mathrm{U}(1)$ gauge field.
The former couples to the $\mathrm{SU}(2)$ spin rotation, while the
latter is the usual electric-magnetic field. The integration of massive fermions
between the D$p$-brane and D$q$-$\overline{\mbox{D}}q$ system
produces the double Chern-Simons coupling,
which agrees with the previously proposed description of quantum spin Hall effect.
One can view
these gauge-interacting TIs/TSCs from D$p$-D$q$-systems
as an analogue of
the projective (parton) construction of the (fractional) QHE
\cite{Wen91-99}.

Our string theory realization of TIs/TSCs
sheds light on extending the projective construction of
the QHE to more generic TIs/TSCs;
it tells us what type of gauge field is ``natural'' to couple with fermions
in topological phases, and guarantees the topological stability
of the system.

This paper is organized as follows.
In Sec.\ \ref{D-brane Construction for Complex Case},
we will present our D-brane construction of TIs/TSCs
for two ``complex'' symmetry classes A and AIII,
which correspond to the complex K-group.
Expecting readers from both high-energy and condensed matter communities,
the underlying principles of the construction are summarized
in Sec.\ \ref{Basic Strategy},
together with a brief but pedagogical review of
D-branes and open strings.
In Sec.\ \ref{D-brane Construction for Real Case},
we will give D-brane
constructions for remaining eight classes which are classified by the real K-group.
This will be followed by a
detailed comparison between the D-brane system and
topological phases in condensed matter physics in
Sec.\ \ref{field theory content}.
In section \ref{Conclusions},
we will draw conclusions and discuss future problems.
In Appendix \ref{AP:spectrum},
we gave a brief
explanation on
the open string spectrum in the presence of D-branes and orientifolds.
This paper is an extend version of our previous brief report
\cite{DbraneShort2010}.

\begin{table}
\begin{center}
\begin{ruledtabular}
\begin{tabular}{c|cccccccc|ccc}
   $\mathrm{class} \backslash d$  & 0 & 1 & 2 & 3 & 4 & 5 & 6 & 7 & T & C & S  \\  \hline
  A & $\mathbb{Z}$ & 0 & $\mathbb{Z}$ & 0 & $\mathbb{Z}$ & 0 & $\mathbb{Z}$ & 0             & 0 & 0 & 0    \\
  AIII & 0 & $\mathbb{Z}$ & 0 & $\mathbb{Z}$ & 0 & $\mathbb{Z}$ & 0 & $\mathbb{Z}$          & 0 & 0 & 1    \\  \hline
  AI & $\mathbb{Z}$ & 0 & 0 & 0 & $2\mathbb{Z}$ & 0 & $\mathbb{Z}_2$ & $\mathbb{Z}_2$    & $+$ & 0 & 0     \\
  BDI & $\mathbb{Z}_2$ & $\mathbb{Z}$ & 0 & 0 & 0 & $2\mathbb{Z}$ & 0 & $\mathbb{Z}_2$     & $+$ & $+$ & 1    \\
  D & $\mathbb{Z}_2$ & $\mathbb{Z}_2$ & $\mathbb{Z}$ & 0 & 0 & 0 & $2\mathbb{Z}$ & 0     & 0 & $+$ & 0     \\
  DIII & 0 & $\mathbb{Z}_2$ & $\mathbb{Z}_2$ & $\mathbb{Z}$ & 0 & 0 & 0 & $2\mathbb{Z}$  & $-$ & $+$ & 1     \\
  AII & $2\mathbb{Z}$ & 0 & $\mathbb{Z}_2$ & $\mathbb{Z}_2$ & $\mathbb{Z}$ & 0 & 0 & 0   & $-$ & 0 & 0     \\
  CII & 0 & $2\mathbb{Z}$ & 0 & $\mathbb{Z}_2$ & $\mathbb{Z}_2$ & $\mathbb{Z}$ & 0 & 0   & $-$ & $-$ & 1     \\
  C & 0 & 0 & $2\mathbb{Z}$ & 0 & $\mathbb{Z}_2$ & $\mathbb{Z}_2$ & $\mathbb{Z}$ & 0     & 0 & $-$ & 0     \\
  CI & 0 & 0 & 0 & $2\mathbb{Z}$ & 0 & $\mathbb{Z}_2$ & $\mathbb{Z}_2$ & $\mathbb{Z}$    & $+$ & $-$ & 1
   \\
\end{tabular}
\end{ruledtabular}
\end{center}
\caption{
\label{charge}
Classification of topological insulators and superconductors
\cite{SRFL,Kitaev};
$d$ is the space dimension;
the left-most column (A, AIII, $\ldots$, CI)
denotes the ten symmetry classes of fermionic Hamiltonians,
which are characterized by the presence/absence of
time-reversal (T), particle-hole (C), and chiral (or sublattice) (S) symmetries
of different types denoted by $\pm 1$ in the right most three columns.
The entries ``$\mathbb{Z}$'', ``$\mathbb{Z}_2$'', ``$2\mathbb{Z}$'',
and ``$0$'' represent the presence/absence of
topological insulators and superconductors,
and when they exist, types of these states
(see Ref.\ \cite{SRFL} for detailed descriptions).
}
\end{table}

\begin{table}[ht]
\begin{center}
\begin{ruledtabular}
\begin{tabular}{cc}
Cartan label
& Riemannian Symmetric Spaces
\\ \hline \hline
A &
$\mathrm{U}(N)\times \mathrm{U}(N)/\mathrm{U}(N)$
\\
AIII &
$\mathrm{U}(N+M)/\mathrm{U}(N)\times \mathrm{U}(M)$
\\
\hline \hline
AI &
$\mathrm{U}(N)/\mathrm{O}(N)$
\\
BDI
&
$\mathrm{O}(N+M)/\mathrm{O}(N)\times \mathrm{O}(M)$
\\
D &
$\mathrm{O}(N)\times \mathrm{O}(N)/\mathrm{O}(N)$
\\
DIII &
$\mathrm{SO}(2N)/\mathrm{U}(N)$
\\
AII &
$\mathrm{U}(2N)/\mathrm{Sp}(2N)$
\\
CII &
$\mathrm{Sp}(N+M)/\mathrm{Sp}(N)\times \mathrm{Sp}(M)$
\\
C &
$\mathrm{Sp}(2N)\times \mathrm{Sp}(2N)/\mathrm{Sp}(2N)$
\\
CI &
$\mathrm{Sp}(2N)/\mathrm{U}(N)$
\\
\end{tabular}
\end{ruledtabular}
\end{center}
\caption{
Cartan classification of Riemannian symmetric spaces
\cite{Helgason1978}.
\label{cartan}
}
\end{table}

\begin{table}
\begin{center}
\begin{ruledtabular}
\begin{tabular}{cccccccccccc}
    & D($-1$) & D0 & D1 & D2 & D3 & D4 & D5 & D6 & D7 & D8 & D9 \\ \hline
  type IIB & $\mathbb{Z}$ & 0 & $\mathbb{Z}$ & 0 & $\mathbb{Z}$ & 0 & $\mathbb{Z}$ & 0 & $\mathbb{Z}$ & 0 & $\mathbb{Z}$ \\
  O9$^{-}$ (type I) & $\mathbb{Z}_2$ & $\mathbb{Z}_2$ & $\mathbb{Z}$ & 0 & 0 & 0 & $\mathbb{Z}$ & 0 & $\mathbb{Z}_2$ & $\mathbb{Z}_2$ & $\mathbb{Z}$ \\
  O9$^{+}$ & 0 & 0 & $\mathbb{Z}$ & 0 & $\mathbb{Z}_2$ & $\mathbb{Z}_2$ & $\mathbb{Z}$ & 0 & 0 & 0 & $\mathbb{Z}$ \\
\end{tabular}
\end{ruledtabular}
\end{center}
\caption{\label{dbrane}
D$p$-brane charges from K-theory, classified by
$\mathrm{K}(\mathbb{S}^{9-p})$,
$\mathrm{KO}(\mathbb{S}^{9-p})$ and
$\mathrm{KSp}(\mathbb{S}^{9-p})$ \cite{WittenK}.
A $\mathbb{Z}_2$ charged D$p$-brane with
$p$ even or $p$ odd represents
a non-BPS D$p$-brane
or a bound state of a D$p$ and an anti-D$p$ brane, respectively \cite{Sen}.
}
\end{table}

\begin{table}
\begin{center}
\begin{ruledtabular}
\begin{tabular}{c|c|cccccccc}
 $G$  &  $\mathrm{class}\backslash d$  & 0 & 1 & 2 & 3 & 4 & 5 & 6 & 7    \\  \hline
U & A   & U & - & U & - & U & - & U & -    \\
U & AIII& - & U & - & U & - & U & - & U     \\  \hline
O & AI  & O & - & - & - &Sp & - & U & O       \\
O & BDI & O & O & - & - & - & Sp& - & U      \\
O & D   & U & O & O & - & - & - & Sp& -       \\
O & DIII& - & U & O & O & - & - & - & Sp      \\
Sp& AII &Sp & - & U & O & O & - & - & -       \\
Sp& CII & - &Sp & - & U & O & O & - & -       \\
Sp& C   & - & - &Sp & - & U & O & O & -       \\
Sp& CI  & - & - & - &Sp & - & U & O & O
   \\
\end{tabular}
\end{ruledtabular}
\end{center}
\caption{\label{gauge}
External $G$ (left-most column) and internal $\tilde{G}$ gauge groups
for each spatial dimension $d$ and symmetry class;
U, O, Sp, represents U(1), $\mathrm{O}(1)=\mathbb{Z}_2$,
and $\mathrm{Sp}(1)=\mathrm{SU}(2)$,
respectively.
}
\end{table}

\section{D-brane Construction for Complex Case}
\label{D-brane Construction for Complex Case}

\subsection{Basic Strategy}
\label{Basic Strategy}

\subsubsection{open string and D-branes}
To realize a TI/TSC in string theory,
we need massive fermions, which are charged under some gauge field.
One of the simplest setups in string theory which gives rise to
charged massive fermions
is a system that consists of two kinds of D-branes: D$p$- and D$q$-branes.

A D$p$-brane is a solitonic object in string theory
whose world volume is $(p+1)$-dimensional.
For our purposes,
a D$p$-brane can simply be regarded as
a $(p+1)$-dimensional plane (for more details refer to,
for example, Ref.\ \cite{Pol}).
An open string can ends at a D-brane,
i.e., a D-brane is defined as the set of end points of open strings.
In other words, an open string attached to
a D$p$-brane satisfies the Neumann boundary condition
in the $(p+1)$ directions
parallel with the D-brane,
while it satisfies the Dirichlet boundary condition
in the remaining $(9-p)$ directions
orthogonal to the D-brane.

An open string has two end points,
and in the presence of $N$ coincident D$p$-branes, say,
these two ends can be attached to
any one of the $N$ D$p$-branes.
Open string states can thus be labeled,
in addition to momentum and
Fock-space states labeling the internal degrees of freedom of the string,
by an entry of an $N\times N$ matrix
(the Chan-Paton matrix).
Indeed we can show
a $(p+1)$-dimensional $\mathrm{U}(N)$ gauge theory lives
on the $N$ coincident D$p$-branes.
In our construction,
the gauge field on the D$p$-brane and D$q$-brane
are denoted by $A_\mu$ and $\ti{A}_\mu$,
which are called the external
and internal gauge field, respectively. See (a) in Fig.\ref{edgestate}.

Open string excitations between the D$p$- and D$q$-branes
give rise to fermions.
(In addition, scalar bosons, which is not our main focus in this paper,
appear sometime as we explain below.)
The mass $m$ of these fermions are proportional to the distance $\Delta x$
between the D$p$- and D$q$-branes,
given by $m=T_{\mathrm{string}}\Delta x$,
where $T_{\mathrm{string}}$
is the string tension (often denoted by $T_{\mathrm{string}}\equiv 1/(2\pi\ap ')$).
We define the common dimension of
the D$p$- and D$q$-branes to be $d+1$ and the
TIs/TSCs in our construction live in the $d+1$ dimensions
because the massive fermions are confined in this directions.

One of the reasons why our strategy to construct
topological phases is based on realizing
charged massive fermions is that,
in the presence of a certain defect (boundary),
quite generically,
such fermions give rise to modes localized at the defect (boundary)
-- a defining property of TIs/TSCs.
(In reality, the appearance of such a boundary is
to some extent inevitable when we
attach probes to measure various physical quantities.)
These points will further be elaborated in Sec.\ \ref{edgest}.

\subsubsection{
Neumann-Neumann,
Dirichlet-Dirichlet,
and
Neumann-Dirichlet directions}

The most important characteristics of
such D-brane systems are the boundary conditions of
open strings stretching between
the D$p$- and D$q$-branes,
as they are the origin of the charged fermions.
Since an open string has two end points each of
which satisfies either
the Neumann (N) or Dirichlet (D)
boundary condition,
we distinguish three possibilities,
i.e.,
(i) both ends satisfy the N boundary condition (NN),
(ii) both ends satisfy the D boundary condition (DD),
and
(iii) mixed boundary condition where one end satisfies the N boundary condition while
the other satisfies the D boundary condition (ND).
The number of these directions,
denoted by $\#\mathrm{NN}$, $\#\mathrm{DD}$ and $\#\mathrm{ND}$,
are completely fixed once we
determine the directions in which the D-branes extend.
Since the total spacetime dimensions
of superstring (called critical dimension) is ten,
they are subjected to the constraint,
\be
\#\mathrm{NN}+\#\mathrm{DD}+\#\mathrm{ND}=10.
\ee

In our construction,
$\#\mathrm{NN}$ and $\#\mathrm{DD}$ have a physical meaning;
$\#\mathrm{NN}$ is equal to the spacetime dimensions of
TIs/TSCs, $\#\mathrm{NN}=d+1$, while
$\#\mathrm{DD}$ represents the number of possible mass deformations
(types of mass perturbations which can be added to the charge fermion system).

In string theory, there is a very important symmetry called T-duality.
If we take a T-duality in a NN or DD direction,
the dimension $d$ decreases or increases since
T-duality turns a NN (DD) direction into a DD (NN) direction,
respectively.
At the same time, type IIA and IIB string theory
are exchanged with each other.
On the other hand, T-duality in a ND direction
separately preserves $\#\mathrm{NN}$, $\#\mathrm{DD}$ and $\#\mathrm{ND}$.
This does not change the dimensions $d$ of TIs/TSCs,
while $p$ and $q$ are shifted.
Therefore, this degree of freedom, taking T-duality in ND directions,
can be thought as somewhat redundant for our purposes,
and we can make use of it to fix the value of $p$.

Once we fix $p$ (we choose $p=5$ for the most part of this section),
each brane system is completely
classified by the two integers $\#\mathrm{NN}=d+1$ and $\#\mathrm{DD}$.
They encode the information
how the D$q$-brane is embedding into the ten dimensional spacetime.
In this picture,
the D$q$-brane charge is identified with the topological charges of the TIs/TSCs.
Indeed, both are classified by K-theory.

Properties of the system realized in the string spectrum
for an open string stretching between the D$p$- and D$q$-branes
depend sensitively on $\#\mathrm{NN}$, $\#\mathrm{DD}$ and $\#\mathrm{ND}$.
For example,
when $\#\mathrm{ND}\leq 4$, the D$p$-D$q$ open string includes scalar bosons
in addition to a massive fermion.
Moreover, when $\#\mathrm{ND}\leq 3$, they become tachyons
which makes the brane system unstable;
this constraints the possible values of
$\#\mathrm{NN}$, $\#\mathrm{DD}$ and $\#\mathrm{ND}$
to realize TIs/TSCs.

Furthermore, we will focus on the cases with $\#\mathrm{DD}=1,2$,
as these cases lead to a stable gapless mode, once we introduce a boundary to the system.
Studying the open string spectrum
between the D$p$- and D$q$-branes for $\#\mathrm{DD}\geq 3$,
we find a stable (=tachyon free) configuration
if $\#\mathrm{DD}\leq 5-d$.
Thus it is possible to consider brane systems with $\#\mathrm{DD}=3,4$
in order to realize TIs/TSCs in $d=1,2$.
However,
since these examples have more $\#\mathrm{DD}$ directions
as compared with the realizations with $\#\mathrm{DD}=1,2$,
they can simply be regarded as multiple copies of
class A or AIII TIs/TSCs.
In other words,
these new configurations have the non-abelian global symmetry
given by $\mathrm{SO}(\#\mathrm{DD})$,
and if we mod out the system by this symmetry,
it reduces to the previous cases.
Therefore we will not discuss examples with $\#\mathrm{DD}\geq 3$
in this paper.

The strategy explained above
is enough to identify D-brane configurations for the class A and AIII,
which belong to the complex category in a K-theoretic sense.
To find D-brane systems for other eight classes, which correspond to the real category, we need
another object called orientifolds (see (b) in Fig.\ref{edgestate}),
which will be discussed separately in the next section.

\begin{figure}[ht]
 \begin{center}
 \includegraphics[height=7cm,clip]{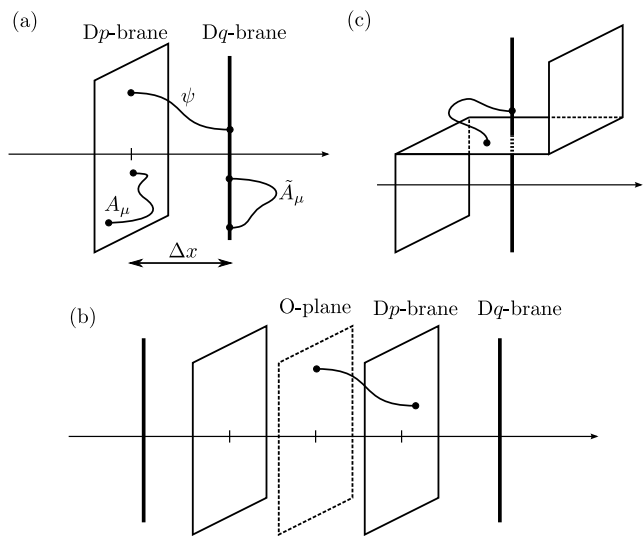}
 \end{center}
 \caption{
 \label{edgestate}
(a) The D$p$-D$q$ system realizing TIs/TSCs in the complex case.
(b) The D$p$-D$q$ system with O-plane realizing TIs/TSCs
in the real case.
(c) The intersecting D$p$-D$q$ system realizing a boundary in TIs/TSCs.
 }
 \end{figure}

\subsection{Class A}

describe the D-brane construction of class A TIs.
In the D-brane construction outlined above,
the system is composed of a D5- and D$q$-brane
($p$ is fixed to be $p=5$ by making use of T-duality).
As mentioned earlier,
we require that the number of mass deformations is one, $\#\mathrm{DD}=1$,
since TIs/TSCs should by definition have gapless modes at the boundary
(edge, surface, etc.).
In our D-brane construction,
as we will come back later in more detail in the section \ref{edgest},
putting a boundary in a TI/TSC is
realized by bending the D$p$(=D5)-brane at the boundary
so that it intersects with the D$q$-brane in an orthogonal way,
as depicted in Fig.\ \ref{edgestate} (c).
Massless modes appear only if
the bent D5-brane intersects with D$q$-brane along the
$d-1$ dimensional boundary.
This is only possible when $\#\mathrm{DD}=1$ under
generic situations.

In this case, we can take a D5-brane in type IIB string theory which extends in the
$x^{0,1,2,3,4,5}$ directions in ten-dimensional space-time without losing
any generality.
The massive fermions live on the $d=q-3$ spacial dimensions which are common to
the D$5$- and D$q$-branes.
We identify these configurations with the class A TIs
in spatial dimensions $d=0,2,4$, as summarized in Table \ref{AandAIII}.
Indeed this
is consistent with the known topological charge in Table \ref{charge}.

Among these, the most familiar example is the QHE (class A in $d=2$).
In our construction, this is
realized as a D5-D5 system. By T-duality, this setup is equivalent to
the D3-D7 system studied in Refs.\ \cite{Rey,DKS,FujitaFQHE}.
Since $\#\mathrm{ND}=6$,
open string excitations between the D5-branes give rise to
two Majorana fermions (Mj) or
equivalently one two-component Dirac fermion (Di), $\psi$,
and no bosons.
For details of counting of fermions, refer to the appendix \ref{AP:spectrum}.
The distance between the D-branes in $x^9$ direction ($\Delta x^9$)
is proportional
to the mass $m$ of the fermions.
The massive fermion is interpreted as an electron and
it couples to an electric-magnetic field $A_\mu$, which is the
$\mathrm{U}(1)$ gauge field on
the D$p$(=D5)-brane.

The low-energy effective theory is schematically summarized by the
effective Lagrangian in the $(2+1)$-dimensional common direction of the two D5-branes,
\begin{eqnarray}
\mathcal{L}
=
\bar{\psi}
[
\gamma^{\mu}
(
i\partial_{\mu} - A_{\mu} -\tilde{A}_{\mu}
)
- m
]
\psi
+
\cdots,
\end{eqnarray}
where the gamma matrices are given,
in terms of the $2\times 2$ Pauli matrices,
by $\gamma^{\mu=0,1,2}=\sigma_3, {i}\sigma_2, -{i}\sigma_1$,
say.
(We will use $\sigma_0$ to denote the $2\times 2$ identity matrix.)

Integrating out the massive fermions yields
the CS term
\be
S_{\mathrm{CS}}
=\frac{k}{8\pi}\int A\wedge dA,
\quad
k = \f{m}{|m|}=\pm 1
\label{CSQHE}
\ee
where we set $\ti{A}=0$ for simplicity.
The Hall conductivity is read off from the CS term for $A_{\mu}$
as
\be
\sigma_{xy}=\f{k}{4\pi}.
\ee
When we change the sign of $m$
by passing the D$q$-brane through the D$p$-brane,
the value of $k$ jumps from $-1$ to $+1$.
If we instead put $N_f$ D$q$-branes,
we have $N_f$ copies of massive Dirac fermions $\psi_i$
which couple with
$\mathrm{U}(1)$ photon field $A_\mu$ and
$\mathrm{U}(N_f)$ gauge fields
$\tilde{A}_{\mu}$ (when all D$q$-branes are coincident), leading to
the Hall conductivity $\sigma_{xy}= N_fk/(4\pi)$.

It is indeed possible to derive the CS term (\ref{CSQHE}) from
the Wess-Zumino (WZ) term of the D5-brane, which expresses how D-branes couple to
various Ramond-Ramond (RR) fields.
Notice that the low energy effective
theory of a D-brane is given by the sum of the Dirac-Born-Infeld action
and the WZ term.
The former is
a non-linear analogue of the Yang-Mills action and is irrelevant in our topological argument below.
Since this $d=2$ example can be naturally
generalized to the $d=0,4$ cases, below we consider the WZ term of the D5-brane
for general dimensions $d$:
\be
S_{\mathrm{WZ}}=
\mu_{\mathrm{D}5}
\int_{\mathrm{D}5}C_{4-d}\we e^{2\pi\al F}, \label{wza}
\ee
where $\mu_{\mathrm{D}5}$ is the RR charge of
the D5-brane and
$C_{n}$ is the RR $n$-form potential and $F=dA$ is the field strength of the
gauge field on the D5-brane.
This is evaluated in the presence of
the D$q$-brane as follows
\be
S_{\mathrm{WZ}}=
\pm\f{1}{2(2\pi)^{d/2}(1+d/2)!}
\int_{\mathbb{R}^{1,d}}\mbox{Tr}[A\we F^{\f{d}{2}}],
\label{cha}
\ee
where we noted the fact that a D$q$-brane induces a half of
the unit RR flux $2\pi\int_{S^{5-d}}dC_{5-d}=\pm 1$
since in our setup the D$p$-brane
is located on either one of the two sides of the D$q$-brane,
but it does not enclose  the D$q$-brane completely.
The sign in Eq.\ (\ref{cha})
depends on which side of the D$q$-brane the D5-brane is situated.
If we set $d=2$ in Eq.\ (\ref{cha}),
we obtain $\pm \f{1}{8\pi}\int A\we F$ and this agrees with
(\ref{CSQHE}). \
Notice also that a topological insulator with
the topological charge $w\in \mathbb{Z}$
is given by the one with $w$ copies
of D$q$-branes and thus has the topological term
given by $w$ times (\ref{cha}).

\begin{figure}[ht]
 \begin{center}
 \includegraphics[height=6.5cm,clip]{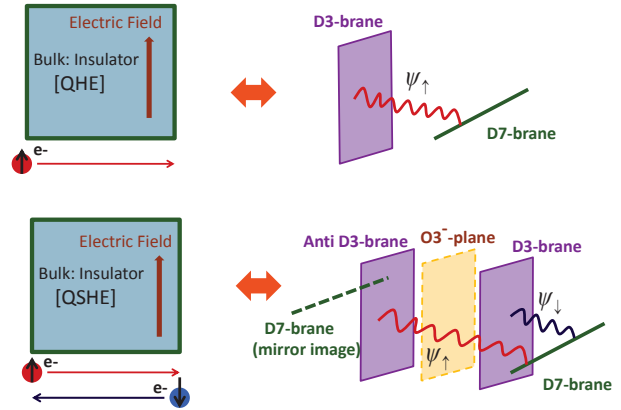}
 \end{center}
 \caption{
Schematic description of the QHE (up) and the QSHE (down)
and their brane configurations (right)
For the presentational convenience,
we took the T-duals of our previous brane setups.
Refer to Table.\ \ref{AIIedge}.
 \label{QHEfig}
 }
 \end{figure}

\subsection{Class AIII}

Now let us turn to class AIII,
which is characterized by the presence of an extra symmetry, SLS.
We argue that SLS is equivalent to an invariance of the brane configurations
under inversion of a coordinate in the Dirichlet-Dirichlet (DD) directions
of open strings between the D$p$ and D$q$-branes.
The existence of boundary gapless modes is guaranteed only if we
assume $\#\mathrm{DD}=2$.
If we take $x^1$- and $x^9$-directions as the two DD directions,
SLS requires $x^1=0$,
which guarantees
the D$p$- and D$q$-branes
intersect when the D$p$-brane is bent
toward the D$q$-brane.

Indeed such a configuration is obtained by taking T-dual
of class A configurations in $x_1$-direction
(Table \ref{AandAIII}).
Again,
the fermion spectrum consists of two Majorana fermions (=one Dirac fermion)
for all dimensions, and
the mass of the fermion is, again, proportional to $\Delta x^9$.
This prescription completely agrees with what is expected from
the Dirac fermion representatives of class AIII TIs in Ref.\ \cite{SRFL};
The $d$-dimensional
class AIII system can be regarded as a dimensional reduction
from the $(d+1)$-dimensional
class A system by setting
one of the components of the momentum to be zero.

In class AIII TIs/TSCs,
the integral topological charge
is related to the coefficient of the topological term $\int F^{(d+1)/2}$.
The coefficient gets shift by one unit
by sign flip $m\to -m$ of the fermion mass $m$.
Consider, as an example,
the $d=3$ case and
assume that there are $N_f$ D6-branes.
By integrating out the $N_f$ massive Dirac fermions,
we find the following topological term for
the $\mathrm{U}(1)$ field strength
\cite{Pavan09}
\be
\f{m}{16\pi^2|m|} N_f\pi \int_{\mathbb{R}^{1,3}} F\wedge F.
\ee
Since we do not assume any time reversal symmetry in this class, we can take
$\theta$ any constant. Its physical consequence is that
if the system has a boundary, there exist $n$ massless
fermions on its boundary.

\begin{table}
\begin{center}
\begin{ruledtabular}
\textbf{Class A:}\\
\begin{tabular}{c|cccccc|cccc|c|cc}
   & $0$ & $1$ & $2$ & $3$ & $4$ & $5$ & $6$ & $7$ & $8$ & $9$ &  $d$ & A \\ \hline
 D5 & $\times$ &  $\times$ & $\times$ & $\times$ & $\times$ &  $\times$ &   &   &   &   &  &   \\ \hline
 D3 & $\times$ &  &  &  &  &        & $\times$ & $\times$ & $\times$ &  &   0 & $\mathbb{Z}$ (2 Mj)  \\
 D5 & $\times$ & $\times$ & $\times$ &  &  &       & $\times$ & $\times$ & $\times$ &   & 2 & $\mathbb{Z}$ (2 Mj)  \\
 D7 & $\times$ & $\times$ & $\times$ & $\times$ & $\times$  &     & $\times$ & $\times$ & $\times$ &   &4 & $\mathbb{Z}$ (1 Di) \\
  \end{tabular}
\end{ruledtabular}
\end{center}
\begin{center}
\textbf{Class AIII:}\\
\begin{ruledtabular}
\begin{tabular}{c|cccccc|ccccc|c|cc}
       & $0$ & $1$ & $2$ & $3$ & $4$ & $5$ & $6$ & $7$ & $8$ & $9$ & & $d$ & AIII  \\ \hline
 D4 & $\times$ &   & $\times$ & $\times$ & $\times$ &  $\times$ &   &   &   &   &    & \\  \hline
 D4 & $\times$ &  & $\times$ &  &  &        & $\times$ & $\times$ & $\times$ &  & & 1 & $\mathbb{Z}$ (2 Mj) \\
 D6 & $\times$ &   & $\times$ & $\times$ & $\times$ &       & $\times$ & $\times$ & $\times$ &  & &3&   $\mathbb{Z}$ (2 Mj) \\
  \end{tabular}
\end{ruledtabular}
\end{center}
\caption{
\label{AandAIII}
D$p$-D$q$ systems for class A and AIII
where $p=5$ and $q=3,5,7$ for A,
and $p=4$ and $q=4,6$ for AIII.
The D-branes extend in the $\mu$-th direction denoted by ``$\times$''
in the ten-dimensional space-time ($\mu=0,\ldots,9$);
$d+1$ is the number of common directions of D$p$-and D$q$-branes;
The last column shows the D$q$-brane charge,
together with fermion spectra per D$q$-brane,
where  "$N_f$ Mj'' or ``$N_f$ Di''
represents $N_f$ flavor of Majorana and Dirac spinors, respectively
\cite{footnote}.}
\end{table}

\subsection{Boundary Gapless Modes of TIs/TSCs}
\label{edgest}

A defining property of TIs/TSCs is the appearance of
stable gapless degrees of freedom,
when the system is terminated by a $(d-1)$-dimensional boundary.
The best-known example is the $(1+1)$-dimensional edge state in the QHE,
i.e. class A in $d=2$.
The integer quantum Hall edge state
is a chiral (Weyl) fermion,
and
always conducting
even in the presence of strong disorder, as far as
the bulk topological character is not destroyed by disorder.

Quite generally,
we can model a boundary in TIs/TSCs
by considering a position dependent fermion mass term.
Let us consider a mass $m=m(y)$ which changes
along one spatial direction, $y$-direction, say,
and changes its sign at $y=0$,
$m(y)>0$ for $y<0$,
$m(y)<0$ for $y>0$,
and
$m(y)=0$ at $y=0$.
I.e., the mass function $m(y)$ has a kink at the boundary
such as $m(y)\propto y$ when $y$ is very small, while $m(y)$ is almost
a constant in the bulk region.

Such kink or domain wall defines
a ($d-1$)-dimensional boundary (or interface)
which is located at $y=0$
and
is flat (straight) along the $(d-1)$ directions.
The band gap vanishes locally at the boundary $m(y=0)=0$.
As well known in the domain wall fermion or
in the index theorems of various forms,
such topological defect (kink)
induces a fermion zero mode localized at the defect,
which in our context is
a gapless fermion mode signaling a topological character of the bulk.

In our brane construction,
the sample boundary can be constructed
by ``bending'' the D$p$-brane toward the D$q$-brane,
to create an intersection between these branes
(Fig.\ \ref{edgestate}).
This leads to a position-dependent fermion mass,
which changes its sign at the intersection as we wanted.
This increases $\#\mathrm{ND}$ by two and the correct number of massless fermions
appears at the intersection.

For example, let us consider the boundary modes for class A and AIII. By bending the D$p$-brane
world-volume in Table \ref{AandAIII} as we mentioned, we find the D-brane configurations with
boundary as described in Table \ref{Aedge}. If we look at class A at $d=1$, then we find a
Weyl fermion and this corresponds to the chiral fermion of the edge state in
the QHE.

\begin{table}
\begin{center}
\begin{ruledtabular}
\textbf{Class A (Boundary):}\\
\begin{tabular}{c|cccccc|cccc|c|cc}
   & $0$ & $1$ & $2$ & $3$ & $4$ & $5$ & $6$ & $7$ & $8$ & $9$ &  $d$ & A \\ \hline
 D5 & $\times$ & $\times$   & & $\times$ & $\times$ & $\times$  &   &   &   & $\times$  &  &   \\ \hline
 D5 & $\times$ & $\times$ & $\times$ &  &  &       & $\times$ & $\times$ & $\times$ &   & 1 & $\mathbb{Z}$ (1 Weyl)  \\
 D7 & $\times$ & $\times$ & $\times$ & $\times$ & $\times$  &     & $\times$ & $\times$ & $\times$ &   & 3 & $\mathbb{Z}$ (1 Weyl) \\
  \end{tabular}
\end{ruledtabular}
\end{center}
\begin{center}
\textbf{Class AIII (Boundary):}\\
\begin{ruledtabular}
\begin{tabular}{c|cccccc|ccccc|c|cc}
       & $0$ & $1$ & $2$ & $3$ & $4$ & $5$ & $6$ & $7$ & $8$ & $9$ & & $d$ & AIII  \\ \hline
 D4 & $\times$ &   &  & $\times$ & $\times$ &  $\times$ &   &   &   & $\times$  &    & \\  \hline
 D4 & $\times$ &  & $\times$ &  &  &        & $\times$ & $\times$ & $\times$ &  & & 0 & $\mathbb{Z}$
 (2 Mj) \\
 D6 & $\times$ &   & $\times$ & $\times$ & $\times$ &       & $\times$ & $\times$ & $\times$ &  & &2&   $\mathbb{Z}$ (2 Mj) \\
  \end{tabular}
\end{ruledtabular}
\end{center}
\caption{
\label{Aedge}
Intersecting D$p$-D$q$ systems realizing
a boundary of class A and AIII TIs/TSCs.}
\end{table}

\section{D-brane Construction for Real Case}
\label{D-brane Construction for Real Case}

To realize other eight ``real'' symmetry classes in string theory,
we need to implement either one of TRS or PHS, or both.
To respect PHS,
which is equivalent to charge conjugation invariance,
we require the gauge fields $A_\mu$ and $\tilde{A}_\mu$
are not independent of their complex conjugates.
This is the same as the orientation ($\Omega$) projection of gauge theories
on D-branes in string theory.

To realize TRS, we recall that it can be viewed as a product of PHS and SLS
\cite{SRFL}.
As SLS in string theory is interpreted as an invariance
under a parity transformation in the transverse coordinates of D-branes,
we can interpret TRS as the multiplication of the complex conjugation and the parity
transformation.
This is precisely what is called the orientifold projection in string theory.

In this way, both TRS and PHS
are related to the complex conjugation,
and symmetry classes with either TRS or PHS (or both)
thus belong to the real category from the K-theoretic viewpoint.

\subsection{Orientifold Projection and Basic Strategy}
\label{Orientifolds Projection and Basic Strategy}

Our strategy to find the D-brane configurations
realizing TIs/TSCs in the eight real symmetry classes goes quite parallel to
the complex case studied in the previous section.
We consider cases with as minimal number of DD directions as possible,
either
$\#\mathrm{DD}=1$ or $\#\mathrm{DD}=2$,
depending on the absence or presence of SLS, respectively.
This is so since cases with $\#\mathrm{DD}\geq 3$
are either unstable
or lead to a system which can simply be viewed as
copies of the one with $\#\mathrm{DD}\geq 1,2$,
as explained before.

The new ingredient in the real case
is the presence of specific orientifolds.
An orientifold projection is
a $\mathbb{Z}_2$ action
defined by
the orientation reverse $\Omega$
followed by
the parity transformation of $9-r$ coordinates
$x_i\to -x_i$,
where $r$ runs from $-1$ to $9$.
The fixed point set of the $\mathbb{Z}_2$ action
is a $(r+1)$-dimensional plane and
called an orientifold $r$-plane
($\mathrm{O}r$-plane).
In superstring theory, there are two kinds of $\Omega$ projection:
orthogonal (O) and symplectic (Sp) types.
They are distinguished by the different actions on the
$N\times N$ matrix for $N$ coincident D-branes. Refer to
\cite{Pol} or the Appendix \ref{AP:spectrum} of this paper for basics
and especially to \cite{Berg} for the analysis of orientifolds in unstable
D-brane systems. Accordingly, there are two kinds of
orientifold plane
called
$\mathrm{O}r^-$- (O) and
$\mathrm{O}r^+$-plane (Sp).
When acting on the gauge fields on $N$ D$p$-branes
parallel to the orientifold plane,
the orientifold projection
projects
the original $\mathrm{U}(N)$ gauge group
into the $\mathrm{SO}(N)$ and $\mathrm{Sp}(N)$ gauge group, respectively.
The T-duality
in the direction parallel to (orthogonal to)
an $\mathrm{O} r^\pm$-plane
leads to
an $\mathrm{O}(r-1)^\pm$
[$\mathrm{O}(r+1)^\pm$]-plane, respectively.

Once we choose either $\mathrm{O}r^-$- or $\mathrm{O}r^+$-plane ,
we then specify D$p$-brane.
We assume that the D$p$-brane has always an integer K-theory topological charge.
In other words, the D$p$-brane preserves supersymmetry in the presence of
$\mathrm{O}r$-plane, though it is broken in general when we add a D$q$-brane later.
On the other hand, for the D$q$-brane,
we allow both supersymmetric and non-supersymmetric configurations.
The D$q$-brane charge is interpreted
as the topological charges of TIs/TSCs,
which can be either $\mathbb{Z}$ or $\mathbb{Z}_2$,
while the D$p$-brane just provides an external gauge field
which measures the topological charges (or Berry phases) induced by the D$q$-brane.
As opposed to the complex case,
the `D$q$-brane' is not always the standard D$q$-brane with GSO
projection of open strings.

As an example,
consider D-branes in type IIB string with
an $\mathrm{O}9^-$-plane, i.e.
type I string theory \cite{WittenK}.
The D-brane charges are summarized in the Table \ref{dbrane}.
The D1, D5 and D9-branes all have the
integer charge $\mathbb{Z}$ and they are the standard BPS D-branes.
On the other hand, D$(-1)$ and D$7$-brane
are $\mathbb{Z}_2$-charged and
they are interpreted as a brane-antibrane system,
i.e.,
D$(-1)$-$\overline{\mbox{D}}(-1)$
and
D$7$-$\overline{\mbox{D}}7$, respectively.
An anti D$p$-brane ($\overline{\mbox{D}}p$) carries
the opposite RR charge as compared with a D$p$-brane
and can be regarded as a higher dimensional
generalization of anti-particle.
Notice that brane-antibrane systems in type II string theory include
tachyons in the open string spectrum
\cite{antibrane}.
However, in the presence of orientation
projection $\Omega$, the tachyons are projected out
and these brane-antibrane systems are stable.
If we consider two copies of such a brane-antibrane system,
then the tachyon appears from the
open string which connects between two different systems \cite{SenSO}.
Therefore the D-brane charge is given by $\mathbb{Z}_2$.
Moreover, one may notice the D$0$ and D$8$-brane are also $\mathbb{Z}_2$ charged.
These are described by non-standard D-branes called non-BPS D-branes \cite{nonBPS},
which are defined by not requiring any GSO projections on the open strings
attached on them.
Note that only odd integer $p$ is allowed for D$p$-branes in type IIB string theory.
Again the tachyons which appear in the absence of the GSO projection,
are eliminated by the $\Omega$ projection.
Similar results can be understood for the
$\mathrm{O}9^+$ by shifting $p\to p+4$
and the results for other values of $p$ can be
obtained by the T-duality transformation.

We can realize all eight symmetry classes
by considering D$p$-D$q$ systems
in the presence of $\mathrm{O}9^\pm$ or $\mathrm{O}8^\pm$
[see Fig.\ \ref{edgestate} (b)]
as we will explain below.
Though it is also possible to proceed to the cases with $\mathrm{O}p^\pm$-planes,
in such systems we often encounter tachyonic instability
due to the open strings between D$p$- and D$q$-branes
as we will discuss later.

\subsection{Class C and D}

First consider the case with O$9^-$ or O$9^+$ by requiring $\#\mathrm{DD}=1$, which guarantees
the existence of gapless modes at the boundary. While only O$9^-$ leads to supersymmetric type I string theory,
here we consider both because the T-dual of O9$^+$
is equivalent to O$p^+$ planes ($p\leq 8$) in type II string theory.
We can take $p=5$ again and can identify
the number of NN directions as spacetime dimensions of TIs/TSCs,
$\#\mathrm{NN}=d+1$.
This leads to the D$5$-D$q$ systems shown in Table \ref{real case}.
We can identify
these configurations
as class C and D,
which are characterized by the presence of PHS.
Notice that these brane configurations
are precisely the $\Omega$ projection of the previous
class A brane configurations,
and
$\Omega$ projection directly corresponds to imposing PHS.
Moreover, we can confirm that the D-brane charges agree with the topological charges of TIs/TSCs
correctly,
and that the fermion contents of these string theory realizations
(denoted in the last two columns in Table \ref{real case}
either by ``$n_f$ Mj'' or ``$n_f$ Di'' with $n_f$ an integer)
perfectly agree with
the Dirac representative of TIs/TSCs constructed in Ref.\ \cite{SRFL}.
The gauge group on D$5$- and D$q$-branes are also $\Omega$ projected and the result is summarized as in
Table \ref{gauge}.

For example, let us consider the $d=2$ case.
The class C and D TIs/TSCs
are described by Chern-Simons theories with the Sp and O
gauge groups, respectively.
If we restrict to the case with the minimum number of D$q$-branes,
then the gauge group on the D$5$- and D$q$-branes
are the same
and it is $\mathrm{SU}(2)$ for class C,
while it is $\mathrm{O}(1)=\mathbb{Z}_2$ for class D.
We may view class C and D TIs/TSCs in $d=2$
as a close cousin of the QHE (class A),
in a sense that the underlying topological field theories
for these TIs/TSCs are the Chern-Simons theories
with gauge group Sp, O, and U, respectively.

\subsection{Class AII and AI}

In order to realize class AII and AI,
we need to impose TRS but not PHS.
As we have explained,
TRS corresponds to the orientifold projection or equally the combination of the $\Omega$ projection and parity. Thus we introduce O$8^\pm$ instead of O$9^\pm$ and set $p=4$. By requiring $\#\mathrm{DD}=1$ as there is no SLS in AII or AI, we obtain the D-brane configurations as shown in
Table \ref{real case} (see also Fig.\ref{QHEfig} for class AII in $d=2$). As is clear from their D-brane charge and the fermion spectrum, we can indeed identify these configurations with the class AII and AI.
Though strictly speaking, the D-brane charges with
an O$p$-plane for $p\leq 8$ are classified by KR-theory, the same result
can be obtained from KO-theory via T-duality for our purpose \cite{ISS}.

\subsection{Class CII, BDI, CI and DIII}

The other four classes possess SLS and thus we need to require $\#\mathrm{DD}=2$ for the
brane constructions. Thus we can obtain their brane systems from the previous C, D, AII and AI
brane setups by shifting $\#\mathrm{DD}$ in an obvious way. They are summarized in
Table \ref{real case} and again they reproduces the correct topological charges and the fermion
spectra. In our convention, the SLS is imposed by requiring $x^9=0$ for all of these classes.

\subsection{Comments}

Even though the ten-dimensional string theories in the bulk are supersymmetric,
our brane setups are not in general.
When $\#\mathrm{ND}=4,8$ with $\mathbb{Z}$ charge,
they exceptionally preserve a quarter of supersymmetries.
In general, when $\#\mathrm{ND}=4$, there exist massive bosons
in addition to the massive fermions.
This happens when $d=4$ for class A, C, AI, AII,
and when $d=3$ for class AIII, CII, CI and DIII.
Since $\#\mathrm{ND}>4$ for all the other brane systems,
we only have fermions from open strings between the D$p$- and D$q$-branes
and there are no tachyons.

In principle, we can take the T-duality further in NN directions.
This generates D$p$-D$q$ systems in the presence of
$\mathrm{O}r^{\pm}$-plane
for $r\leq 7$ as in Table \ref{gen}.
This seems to lead to theories with different properties than the ten classes of TIs/TSCs.
However, they have $\#\mathrm{DD}\geq 3$
and as we explained before, such systems are either
unstable because of open string tachyons,
or essentially multiple copies of the ten classes which we already
constructed from D-branes. Thus our D-brane construction reproduces precisely the
ten fold classifications given in \cite{SRFL}.

Note also that we have succeeded to realize all TIs/TSCs
in space dimensions $d\leq 4$,
even though string theory lives in ten dimensions.
This is so since we need to consider the common spacetime of
D$p$- and D$q$-branes.
The existence of boundary gapless modes requires $\#\mathrm{DD}\geq 1$
and the absence of tachyon argues
$\#\mathrm{ND}\geq 4$. Thus we inevitably have the constraint
\be
d=\#\mathrm{NN}-1=9-\#\mathrm{DD}-\#\mathrm{ND}\leq 4.
\ee

\begin{table}
\begin{center}
\textbf{Class C and D:}\\
\begin{ruledtabular}
\begin{tabular}{c|cccccc|cccc|c|c|c}
   & $0$ & $1$ & $2$ & $3$ & $4$ & $5$ & $6$ & $7$ & $8$ & $9$& $d$  & C (O$9^{-}$) & D (O$9^{+}$) \\ \hline
 D5 & $\times$ & $\times$  & $\times$ & $\times$ & $\times$ & $\times$       &  &  &  &  &  &   \\ \hline
 D3 & $\times$ &   &  &  &  &        & $\times$ & $\times$ & $\times$ &   & 0 & 0 & $\mathbb{Z}_2$ (2 Mj)\\
 D4 & $\times$ & $\times$ &   &  &  &       & $\times$ & $\times$ & $\times$ & & 1 & 0 &  $\mathbb{Z}_2$ (1 Mj) \\
 D5 & $\times$ & $\times$ & $\times$ &   &   &     & $\times$ & $\times$ & $\times$ &  & 2 &  $\mathbb{Z}$ (4 Mj)  &  $\mathbb{Z}$ (1 Mj) \\
 D6 & $\times$ & $\times$  & $\times$ & $\times$ &  &        & $\times$ & $\times$ & $\times$ & & 3  & 0 & 0 \\
 D7 & $\times$ & $\times$ & $\times$ & $\times$ & $\times$ &        & $\times$ & $\times$ & $\times$ & & 4  & $\mathbb{Z}_2$ (2 Di) & 0 \\
  \end{tabular}
\end{ruledtabular}
\end{center}
\begin{center}
  \textbf{Class CI and DIII:}\\
\begin{ruledtabular}
\begin{tabular}{c|cccccc|cccc|c|c|c}
   & $0$ & $1$ & $2$ & $3$ & $4$ & $5$ & $6$ & $7$ & $8$ & $9$ & $d$ & CI (O$9^{-}$) & DIII (O$9^{+}$) \\ \hline
   D5 & $\times$ & $\times$  & $\times$ & $\times$ & $\times$ & $\times$       &  &  & & &  & &  \\ \hline
 D2 & $\times$ &   &  &  &  &        & $\times$ & $\times$ &  & &0  & 0 & 0 \\
 D3 & $\times$ & $\times$ &   &  &  &       & $\times$ & $\times$ &  & &1  & 0 &  $\mathbb{Z}_2$ (2 Mj) \\
 D4 & $\times$ & $\times$ & $\times$ &   &   &     & $\times$ & $\times$ &  & &2  & 0 & $\mathbb{Z}_2$ (2 Mj) \\
 D5 & $\times$ & $\times$  & $\times$ & $\times$ &  &        & $\times$ & $\times$ &  & &3   &  $\mathbb{Z}$ (4 Mj)  & $\mathbb{Z}$ (1 Mj)
  \end{tabular}
\end{ruledtabular}
\end{center}
\begin{center}
\textbf{Class AII and AI:}\\
\begin{ruledtabular}
\begin{tabular}{c|ccccc|c|cccc|c|c|cc}
    &$0$& $1$ & $2$ & $3$ & $4$ & $5$ & $6$ & $7$ & $8$ & $9$ & $d$ & AII (O$8^{-}$)  & AI (O$8^{+}$) \\ \hline
D4&$\times$&$\times$&$\times$&$\times$ &$\times$&        &  &  &  &  &  &   \\ \hline
D4&$\times$&            &            &              &             &        &$\times$&$\times$&$\times$&$\times$& 0 & $\mathbb{Z}$ (4 Mj) & $\mathbb{Z}$ (1 Mj)\\
D5&$\times$&$\times$&            &  &  &       & $\times$ & $\times$ & $\times$ & $\times$ & 1 &  0 &  0 \\
D6&$\times$&$\times$&$\times$&   &   &     & $\times$ & $\times$ & $\times$ & $\times$ & 2 &  $\mathbb{Z}_2$ (4 Mj) & 0 \\
D7&$\times$&$\times$&$\times$& $\times$ &   &     & $\times$ & $\times$ & $\times$ & $\times$ & 3 &  $\mathbb{Z}_2$ (2 Mj) & 0 \\
D8&$\times$&$\times$&$\times$& $\times$ & $\times$ &        & $\times$ & $\times$ & $\times$ &  $\times$ & 4 & $\mathbb{Z}$ (1 Di) & $\mathbb{Z}$ (1 Di) \\
  \end{tabular}
\end{ruledtabular}
\end{center}
\begin{center}
\textbf{Class CII and BDI:}\\
\begin{ruledtabular}
\begin{tabular}{c|ccccc|c|cccc|c|c|c}
 & $0$ & $1$ & $2$ & $3$ & $4$ & $5$ & $6$ & $7$ & $8$ & $9$ & $d$ & CII (O$8^{-}$)  & BDI (O$8^{+}$) \\ \hline
D4 & $\times$ & $\times$  & $\times$ & $\times$ & $\times$ &        &  &  &  &  & &   \\ \hline
D3 & $\times$ &   &  &  &  &        & $\times$ & $\times$ & $\times$ &   & 0 & 0 & $\mathbb{Z}_2$ (2 Mj)\\
D4 & $\times$ & $\times$ &   &  &  &       & $\times$ & $\times$ & $\times$ &  & 1 &  $\mathbb{Z}$ (4 Mj) &  $\mathbb{Z}$ (1 Mj) \\
D5 & $\times$ & $\times$ & $\times$ &   &   &     & $\times$ & $\times$ & $\times$ & & 2 &  0 & 0 \\
D6 & $\times$ & $\times$  & $\times$ & $\times$ &  &        & $\times$ & $\times$ & $\times$ &  & 3 & $\mathbb{Z}_2$ (4 Mj) & 0 \\
D7 & $\times$ & $\times$  & $\times$ & $\times$ & $\times$  &        & $\times$ & $\times$ & $\times$ &  & 4 & $\mathbb{Z}_2$ (2 Di) & 0
  \end{tabular}
\end{ruledtabular}
\end{center}
\caption{
D$p$-D$q$ systems for eight ``real'' symmetry classes,
where $p=5$ for classes C, D, CI, DIII,
and $p=4$ for classes AII, AI, CII, BDI.
For classes AII, AI, CII, BDI,
the O8-plane extends except $x^5$
\cite{footnote}.
In these tables, D$p$ with $\mathbb{Z}_2$ charge
denotes
a D$p-\overline{\mbox{D}}p$ system for $p$ odd (or even)
in type IIB (or IIA) string theory,
a non-BPS D-brane for $p$ even (or odd)
in type IIB (or IIA) string theory.
\label{real case}
}
\end{table}

\subsection{Field Theory Content}
\label{field theory content}

We now describe the field theory content of our brane systems
in more details and their link to condensed matter systems.

\subsubsection{primary series}

First, let us focus on the diagonal in Table \ref{charge},
where lie TIs/TSCs characterized by an integer topological invariant.
These TIs/TSCs can be called ``primary series'',
since lower-dimensional
TIs/TSCs characterized by a binary topological invariant
can be derived from them by dimensional reduction.
\cite{Qi2008, SRFL}
For $d=2n$ dimensional TIs/TSCs,
the relevant topological invariant is the Chern invariant
defined for the (non-Abelian) Berry connection in momentum space,
while
for $d=2n+1$ dimensional TIs/TSCs,
the relevant topological invariant is
the Chern-Simons invariant,
and the winding number.

As we have been emphasizing,
our string theory construction of TIs/TSCs
naturally gives rise to gauge interactions,
together with massive fermions.
For the primary series,
the internal gauge group is $\mathrm{O}(1)=\mathbb{Z}_2$.
The $\mathbb{Z}_2$ gauge theory is somewhat difficult to formulate
in the continuum spacetime --
it is better put on a discretized spacetime (lattice).
In string theory, the $\mathbb{Z}_2$ gauge group is visible
when we consider a Wilson loop which is projected
by the $\mathbb{Z}_2$ projection.
The original Wilson loop takes continuous values of $\mathrm{U}(1)$ but after the
orientifold projection it only takes $\mathbb{Z}_2$ values $\pm 1$.

The appearance of the $\mathbb{Z}_2$ gauge group is particularly indicative,
since there exist interacting bosonic models defined on a lattice,
which realize a topological phase
that is described in terms of (emergent) fermions
interacting with a $\mathbb{Z}_2$ gauge field.
The microscopic model was first introduced by Kitaev
\cite{Kitaev05} for $S=1/2$ spins
on the honeycomb lattice, and can be exactly solvable.
Since then, a huge number of generalizations/extensions of the original Kitaev
model has been discussed.
The original Kitaev model realizes,
in the phase with broken TRS,
a topological phase with non-Abelian statistics,
which can be described in terms of
emergent Majorana fermions in symmetry class D
interacting with $\mathbb{Z}_2$ gauge field.
Our string theory realization of
the TI/TSC in class D in $d=2$
can be viewed as corresponding to
the honeycomb lattice Kitaev model in the non-Abelian topological phase.
A three-dimensional
generalization of the honeycomb lattice Kitaev model
to the diamond lattice was discussed in Ref.\ \cite{Ryu08}.
As in the original Kitaev model,
for a region of its phase diagram,
the model realizes a topological phase
which can be described in terms
emergent Majorana fermions in symmetry class DIII
interacting with $\mathbb{Z}_2$ gauge field.
Again, our string theory realization of
the TI/TSC in class DIII in $d=3$
can be viewed as corresponding to
(a proper supersymmetric generalization of)
this interacting bosonic model.
(For AII ($d=4$) and AI ($d=0$), there is 1/4 susy,
and for DIII ($d=3$),
there is 1/8 susy.
For other TIs/SCs in the primary series, there is no susy.)

\subsubsection{first descendant}

Let us now focus on TIs/TSCs
with binary $\mathbb{Z}_2$ topological invariant
lying beneath/immediate left to
the primary series
in Table \ref{charge},
i.e,
BDI ($d=0$), D ($d=1$), DIII ($d=2$), and AII ($d=3$).
These TIs/TSCs can be obtained from the primary series
by dimensional reduction, and therefore can be called
the first descendant.
The internal gauge group is $\mathrm{O}(1)=\mathbb{Z}_2$
(Table \ref{gauge}).
In our string theory realization of TIs/TSCs,
the first descendant is realized by
a D-brane with $\mathbb{Z}_2$ K-theory charge,
which is called a non-BPS D-brane \cite{nonBPS}.

\paragraph{the $\mathbb{Z}_2$ topological insulator
(class $\mathrm{AII}$ in $d=3$)}

Among the $\mathbb{Z}_2$ TIs/TSCs of the first descendent
is the $d=3$-dimensional $\mathbb{Z}_2$ topological insulator in class AII
which is of most interest recently.
The external gauge group for this case is
$G=\mathrm{Sp}(1)=\mathrm{SU}(2)$
(Table \ref{gauge}).
By moving the D$p$-brane away from O8-plane,
this can further be broken into its diagonal subgroup $\mathrm{U}(1)$,
which couples to fermions as an external ``electromagnetic''
$\mathrm{U}(1)$ gauge field.
In this case, after integrating out the fermions,
the $\theta$ term at $\theta=\pi$,
\be
\f{m}{16\pi|m|}
\int_{\mathbb{R}^{1,3}} F\wedge F
\ee
will be induced,
in agreement with Refs.\ \cite{Qi2008, Essin08}.

\subsubsection{second descendant}

The $\mathbb{Z}_2$ TIs/TSCs
lying beneath/immediate left to
the first descendent in Table \ref{charge},
i.e,
D ($d=0$), DIII ($d=1$), AII ($d=2$), and CII ($d=3$),
can be called the second descendant,
since they can be obtained by dimensionally
reducing the primary series twice.
For these $\mathbb{Z}_2$ TIs/TSCs
the internal gauge group is U(1).
In our string theory realization of TIs/TSCs,
the second descendant is realized by
a D-brane with a $\mathbb{Z}_2$ K-theory charge,
which is a bound state of a D-brane and a anti D-brane
(so called brane-antibrane system) \cite{antibrane}.

For class D and DIII,
the U(1) gauge field
couples to two real fermions
which can be combined into a single complex field.
For class AII and CII, the fermion spectrum consists of 4 Mj.
In this case,
the external gauge field is $\mathrm{Sp}(1)=\mathrm{SU}(2)$,
which couples to a doublet, $\psi_{\uparrow/\downarrow}$.
Each has 2 Mj (=1Di) degrees of freedom and
couples to a U(1) internal gauge field.

\paragraph{the quantum spin Hall effect
(class $\mathrm{AII}$ in $d=2$)}

Among $\mathbb{Z}_2$ TIs/TSCs in the second descendants,
we now take a closer look at a TI in
class $\mathrm{AII}$ in $d=2$,
which is the QSHE described in
Fig.\ \ref{QHEfig}.
The low-energy effective theory
of the D$p$-D$q$ system
is summarized by the
following effective Lagrangian
\begin{eqnarray}
\mathcal{L}
=
\bar{\psi}
[
\gamma^{\mu}(
i\partial_{\mu} - A_{\mu} -\tilde{A}_{\mu}
)
- m M
] \psi
+\cdots,
\label{Dirac QSHE}
\end{eqnarray}
where $\psi= (\psi_{\uparrow},\psi_{\downarrow})^T$,
and
$M$ is a diagonal mass matrix
whose eigenvalue is $\pm 1$ for $\psi_{\uparrow/\downarrow}$,
respectively.
The gamma matrices and mass matrices are given,
in terms of two sets of the $2\times 2$ Pauli matrices
$\sigma_{1,2,3}$ and $\tau_{1,2,3}$
by
$\gamma^{\mu=0,1,2}=
\sigma_3\otimes \tau_0,
{i}\sigma_2\otimes \tau_0,
-{i}\sigma_1\otimes \tau_0$,
and
$M=\sigma_0 \otimes \tau_3$,
say.
The $4\times 4$ Dirac Hamiltonian of the same type has been
widely discussed as a model of the QSHE
\cite{KaneMele,bernevig06}
The edge mode of the QSHE consists of
odd number of Kramers pairs, which is just one for the
Dirac model (\ref{Dirac QSHE}).

The gauge group of this theory is
$G\times \tilde{G}
=\mathrm{SU}(2)\times \mathrm{U}(1)$,
i.e., the gauge field
$A_\mu$ on the D$p$-brane is SU(2),
while $\ti{A}_\mu$ on the D$q$-brane is U(1).
The $\mathrm{U}(1)$ gauge field
can be interpreted as the usual electric-magnetic field,
while the $\mathrm{SU}(2)$
can be viewed as a (fictitious) gauge field which couples
to the $\mathrm{SU}(2)$ spin.
Below we only consider the third component
(proportional to $\tau_3$)
of the $\mathrm{SU}(2)$ gauge field $A^3_\mu$.
By integrating out the
bi-fundamental fermions $\psi$, we find
the double Chern-Simons action
($A_{\mu}=A^3_{\mu}$),
\begin{eqnarray}
S_{\mathrm{dCS}}
\!\!&=&\!\!
\f{1}{8\pi}\int (A+\ti{A})\we d(A+\ti{A})\no
&&\ \ \ -\f{1}{8\pi}\int (A-\ti{A})\we d(A-\ti{A})
\nonumber \\
\!\!&=&\!\!
\f{1}{2\pi}\int A\we d\ti{A}.
\end{eqnarray}
From the effective action
we can find
the QSHE,
\be
j^\mu_{\mathrm{spin}}=
\f{1}{2}\f{\delta S_{\mathrm{dCS}}}{\delta A_\mu}
=
\f{1}{4\pi}
\epsilon^{\mu\nu\rho}
\partial_{\nu}
\ti{A}_{\rho},
\ee
where the factor $1/2$
comes
from the basic fact that the spin of an electron is
$\hbar/2$.
(See for example, Refs.\ \cite{QSHEZ,Ryumasses}.)

\subsubsection{TIs/TSCs with $2\mathbb{Z}$ topological charge}

Finally,
for TIs/TSCs labeled by $2\mathbb{Z}$ in Table \ref{charge},
i.e.,
AII ($d=0$), CII ($d=1$), C ($d=2$), and CI ($d=3$),
the gauge group is $G \times \tilde{G} = \mathrm{SU}(2)\times \mathrm{SU}(2)$,
with 4 Mj fermions in bi-fundamental.

An interesting example of
TSC in class C ($d=2$)
is the $d+id$-wave SC,
which is spin singlet with broken TRS
\cite{Read00,SenthilMarstonFisher}.
A lattice model of
three-dimensional topological SC
in class CI,
which is spin singlet
and invariant under time-reversal,
was constructed in Ref.\ \cite{SchnyderRyuLudwig2009}.

\begin{table}[t]
\begin{center}
\begin{ruledtabular}
\begin{tabular}{c|ccccc}
  $\#\mathrm{DD}$ & (O$9^{-}$\!\!,O$9^+$)\!\! & (O$8^{-}$\!\!,O$8^+$)\!\!
  & (O$7^{-}$\!\!,O$7^+$)\!\!
  & (O$6^{-}$\!\!,O$6^+$)\!\! & (O$5^{-}$\!\!,O$5^+$\!\!)\!\!  \\ \hline
  0 & Chiral &  &   &   & \\
  1 & (C,D) &  (AII,AI) &  &  &   \\
  2 & (CI,DIII) & (CII,BDI) & (DIII,CI) &  &  \\
  3 & $d\leq 2$ & $d\leq 2$ & $d\leq 2$ & $d\leq 2$ &    \\
  4 & $d\leq 1$ & $d\leq 1$ & $d\leq 1$ & $d\leq 1$  & $d\leq 1$  \\
  \end{tabular}
\end{ruledtabular}
\end{center}
\caption{
\label{gen}
D$p$-D$q$ systems in the presence of an O-plane;
``Tac'' represents
the presence of tachyons in open strings
between D$p$ and D$q$
at least when $d=3$;
``$d\ge 2$'' means
the constraint of possible spatial dimensions $d$
in the brane systems. This is due to either
an absence of corresponding brane systems or
tachyonic instability. ``Chiral'' denotes the existence of chiral fermions
and
is interpreted as boundary (edge) states.
The horizontal direction is shifted by a T-duality in the NN direction.
}
\end{table}

\begin{table}
\begin{center}
\begin{ruledtabular}
\textbf{Class AII:}\\
\begin{tabular}{c|cccccc|cccc|c|cc}
   & $0$ & $1$ & $2$ & $3$ & $4$ & $5$ & $6$ & $7$ & $8$ & $9$ &  $d$ & AII \\ \hline
$\mathrm{O}3^-$ & $\times$ & $\times$   & $\times$ & $\times$ &  & &   &   &   &   &  &   \\
 D7(=D$p$) & $\times$ & $\times$   & $\times$  & $\times$ &  & $\times$  & $\times$  & $\times$  & $\times$
  &   &  &   \\ \hline
  D1 & $\times$ &  & &  & $\times$  &       &  &  &  &   & 0 & $\mathbb{Z}$ (4 Mj)  \\
 D2 & $\times$ & $\times$ & &  & $\times$  &       &  &  &  &   & 1 &   \\
 D3 & $\times$ & $\times$ & $\times$ &  & $\times$  &       &  &  &  &   & 2 & $\mathbb{Z}_2$ (4 Mj)  \\
 D4 & $\times$ & $\times$ & $\times$ &  $\times$ & $\times$  &       &  &  &  &   & 3
 & $\mathbb{Z}_2$ (2 Mj)  \\
   \end{tabular}
\end{ruledtabular}
\end{center}
\begin{center}
\begin{ruledtabular}
\textbf{Class AII (Boundary):}\\
\begin{tabular}{c|cccccc|cccc|c|cc}
   & $0$ & $1$ & $2$ & $3$ & $4$ & $5$ & $6$ & $7$ & $8$ & $9$ &  $d$ & AII \\ \hline
$\mathrm{O}3^-$ & $\times$ & $\times$   & $\times$ & $\times$ &  & &   &   &   &   &  &   \\
 Edge D7 & $\times$ &    & $\times$  & $\times$ &  & $\times$  & $\times$  & $\times$  & $\times$
  & $\times$  &  &  \\ \hline
 D3 & $\times$ & $\times$ & $\times$ &  & $\times$  &       &  &  &  &   & 1 & $\mathbb{Z}_2$ (2 Mj)  \\
 D4 & $\times$ & $\times$ & $\times$ &  $\times$ & $\times$  &       &  &  &  &   & 2
 & $\mathbb{Z}_2$ (2 Mj)
   \end{tabular}
\end{ruledtabular}
\end{center}
\caption{
\label{AIIedge}
D-brane configurations for class AII TIs
in a T-dualized setup (top) and
the brane construction of their boundary gapless modes (bottom).
Here, D3 and D4 express a D3-$\overline{\mathrm{D}}$3 system
and a non-BPS D4-brane, respectively.}
\end{table}

\subsection{Boundary Gapless Modes of TIs/TSCs}

As in the complex case,
boundary modes of the real case
can be realized as an intersecting D$p$-D$q$ system.
Here let us focus on class AII.
For convenience,
we take T-duality of the setup in Table \ref{real case}
in five ND directions,
and consider the setup in the first table in \ref{AIIedge},
whose sketch is depicted in Fig.\ \ref{QHEfig}.
Intersecting D$p$-D$q$ systems describing the boundary
of class AII TIs can then be obtained by
exchanging $x^1$ and $x^9$ for the D7-brane,
as in the second table in \ref{AIIedge}.

The analysis on the open string spectrum
shows that the edge mode in $d=2$, i.e. the QSHE,
is not chiral as opposed to class A TIs as expected. Indeed, the
D$q$-brane is now given by the D$3$-$\overline{\mbox{D}}3$
brane-anti-brane system
(see Sec.\ \ref{Orientifolds Projection and Basic Strategy})
and its intersection
with the bent D7-brane leads to left and right-moving fermion on the $1+1$ dimensional
boundary (see Fig.\ \ref{edgefig}).

We can also see that the edge state
configurations shown in Table \ref{AIIedge}
are actually unstable because the D7-brane
bent toward the O3-plane
is reflected back as an anti D7-brane at the O-plane.
Since we inevitably have a parallel D7- and $\overline{\mathrm{D}}$7-brane
at the same time,
this system has in general tachyons and is unstable
unless we fine tune its intersection angle.
This actually is nothing but requiring TRS,
since the open string tachyon field $T$ is odd under time-reversal.
This can be regarded as an example where the presence of interactions
non-trivially change stabilities of topological insulators. A similar instability
occurs for class AI, CII and BDI.

\begin{figure}[ht]
 \begin{center}
 \includegraphics[height=5cm,clip]{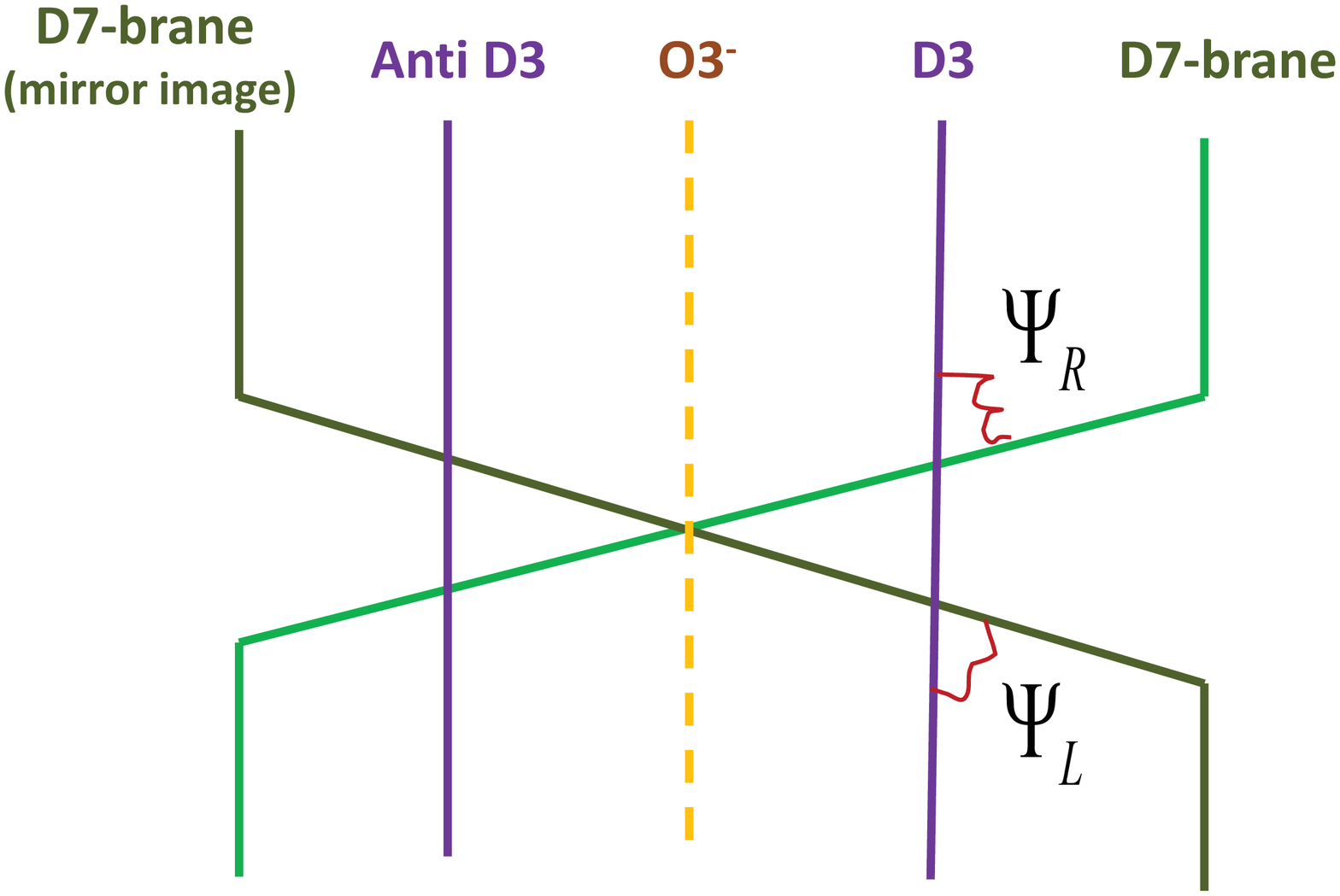}
 \end{center}
 \caption{The brane configuration of the edge state of the QSHE
(class AII in $d=2$).
 \label{edgefig}
 }
 \end{figure}

\section{Conclusions}
\label{Conclusions}

The tenfold classification of TIs/SCs
was derived previously by several different approaches.
One of the reasonings in Ref.\ \cite{SRFL}
that leads to the classification is
the so-called boundary-to-bulk approach;
i.e,.
exhaustive study of non-Anderson localized state
at boundaries which signals the existence of the
topological bulk theory.
On the other hand, Kitaev
proposed K-theory approach for the classification.
Our string theory realization of TIs/TSCs
can be thought of yet another derivation of
the tenfold classification.
It is quite compelling that all different approaches
give the identical table.

It is worth while mentioning that
the construction of TIs/TSCs from D$p$-D$q$ systems
is highly constrained in the following sense:
D$p$-D$q$ systems
which do not correspond
to a bulk TI/TSC or a boundary of a TI/TSC
(i.e., an intersecting D$p$-D$q$ system)
suffer from
the existence of tachyons in the open string spectrum
or are just multiple copies of the minimal fermion
systems for the ten classes.

It is also interesting to note that,
for the Dirac representative of TIs/TSCs
which are constructed systematically in Ref.\ \cite{SRFL},
the \textit{momentum} dependence of
the projection operator of Bloch wavefunctions,
which is one of key ingredients in the classification of TIs/TSCs,
looks quite similar to
the spatial profile of the tachyon field
in string theory
in \textit{real} space \cite{WittenK,Hor}.

One of the implications
of our D-brane construction of TIs/TSCs
is the underlying field theory description.
When a charge of some sort is conserved,
one can describe the topological phase
in terms of effective field theory for the linear responses,
but this is not always the case.
For example,
the $d=2$-dimensional p-wave SC
is one of representative and most important
topological phases in $d=2$,
but its description in terms of field theory
is not yet clearly understood, partly because
there is no conserved quantum number.
An effective (topological) field theory description,
if exists,
suggests the topological phase in question is
robust against several perturbations,
in particular against interactions.
D-branes not only carry a K-theory charge
which is necessary to describe topological phases,
but they also come with the WZ term.
This sheds a light on
field theory description of TIs/TSCs in all symmetry classes.

Finally, to put our D-brane construction in a perspective,
we can consider holographic descriptions of these systems by extending
the constructions in \cite{DKS,FujitaFQHE} in principle,
though it is not possible to take the large-$N$ limit of
$\mathbb{Z}_2$ charged D-branes.
For a closely related holographic construction of
fractional
topological insulators, refer to the recent paper
\cite{HoyosBadajozJensenKarch2010}. It would be interesting to
holographically calculate the entanglement entropy \cite{RTEE}
as it is expected to distinguish various topological phases, including
topological insulators/superconductors \cite{TopEE}.


\acknowledgments
We acknowledge
 ``Quantum Criticality and the AdS/CFT correspondence''
 miniprogram at KITP,
 ``Quantum Theory and Symmetries 6''
 conference at University of Kentucky.
We would like to thank
K.\ Hori, A.\ Karch, A.\ Kitaev, P.\ Kraus, A.\ Ludwig, J.\ Moore,
M.\ Oshikawa, V.\ Schomerus, and S.\ Sugimoto
for useful discussion,
and A. Furusaki for his clear lecture at
``Development of Quantum Field Theory and String Theory'' (YITP-W-09-04)
at Kyoto University.
 SR thanks Center for Condensed Matter Theory at University of
 California, Berkeley for its support.
 TT is supported in part by JSPS Grant-in-Aid
 for Scientific Research No.\ 20740132, and
 by JSPS Grant-in-Aid for Creative Scientific Research No.\ 19GS0219.

\appendix
\section{Overview on the Calculations of Open String Spectra}
\label{AP:spectrum}

In this appendix, we give an overview on how to calculate
the fermion spectrum (the number of fermion species)
arising from open strings
between D$p$- and D$q$-branes.
We neglect all stringy modes and focus
on the low energy limit.
For more details, refer to textbooks, e.g., Ref.\ \cite{Pol}.

\subsection{D-branes in Type II String Theory (Complex Case)}

We first consider unoriented strings and do not impose
the orientifold projection.
If we set $\# \mathrm{ND}=2k$,
then the number $n_F$ of independent components of on-shell
complex fermions in
the D$p$-D$q$ system is given by
\be
n_F=2^{5-k}\times \f{1}{2}\times \f{1}{2}=2^{3-k},
\ee
where each of the three factors describe
the degeneracy of the fermion zero modes in $\mathrm{NN}$ and $\mathrm{DD}$ directions,
the GSO projection,
and the Dirac equation constraint, respectively.
The mass of these fermions are proportional to the distance between the D$p$- and D$q$-branes.
Notice that this counting has already taken
into account the two orientations of the open string.
For example, for the D5-D5 system with $k=3$,
we find $n_F=1$, i.e. one complex fermion in $2+1$ dimensions.
If we consider the D1-D9 system,
we have $k=4$ and thus find $n_F=1/2$, which means one chiral left-moving
fermion in $1+1$ dimensions.

In the case where $\# \mathrm{ND}=2k-1$,
the D$q$-brane is a non-BPS brane
\cite{nonBPS},
which is defined by a D-brane without the GSO projection.
We thus get
\be
n_F=2^{5-k}\times  \f{1}{2}=2^{4-k}.
\ee

\subsection{Orientifold Projections (Real Case)}

We now consider oriented strings to describe the real case,
and take into account
the orientifold projection (denoted by $\Omega$).
The $\Omega$ projection in the csae of the O9-plane
acts on D$p$-branes as follows
\ba
&& \Omega |\mathrm{D}p\lb = |\ov{\mathrm{D}p}\lb,\ \ \ p=-1,3,7 \no
&& \Omega |\mathrm{D}p\lb = |\mathrm{D}p\lb,\ \ \ p\in \mbox{others}
\label{actioncp}
\ea
where $|\mathrm{D}p\lb$ and $|\ov{\mathrm{D}p}\lb$ are the boundary states of
D$p$-brane and anti D$p$-brane, which express the D-brane states in the closed string
Hilbert space.
This is simply because in type I string theory, the 0 and 4-form RR fields are
projected out.
In order to understand the action of $\Omega$
on open string spectrum, we need to introduce the action of $\Omega$
on the Chan-Paton factor.
There are two possible actions: SO (called $\mathrm{O}9^-$-plane)
and Sp (called $\mathrm{O}9^+$-plane) type. Each of them is defined by the action on the Chan-Paton
matrix $\Lambda$:
\be
\Lambda\to \gamma\Lambda^T \gamma^{-1},
\ee
with
\be
\gamma_{\mathrm{SO}}=1,\ \ \ \ \ \ \gamma_{\mathrm{Sp}}=\sigma_2 \otimes 1.
\ee
In addition we need to take into account (\ref{actioncp}).
As clear from the definition, the dimension of the Chan-Paton matrix for the Sp projection has to be even.

For example, for a D1-brane, the massless bosonic modes
(the gauge field $A_\mu$ and the transverse scalars
$\phi^i$)
obey the following projection in the presence of $\mathrm{O}9^-$-plane:
\ba
A_\mu^T=- A_{\mu},
\quad
(\phi^i)^T=\phi^i,
\ea
where the minus sign in front of the gauge field
is due to the $\Omega$ action on the world-sheet oscillators.
This leads to the gauge group $\mathrm{SO}(N)$.
On the other hand, in the presence of $\mathrm{O}9^+$-plane, we have the projection
\be
\gamma_{\mathrm{Sp}}A_\mu^T
\gamma_{\mathrm{Sp}}=-A_\mu,\ \ \ \
\gamma_{\mathrm{Sp}}\phi_i^T\gamma_{\mathrm{Sp}}=\phi_i,
\ee
from which we obtain the gauge group $\mathrm{Sp}(N)$.

On the other hand, if we are interested in the open string between two different D-branes, then the $\Omega$
projection just relates the two open strings extending opposite directions.
If we define $\Psi_{12}$ as a field from such a open string, this projection is explicitly given
as follows
\be
\gamma_{1}\Psi_{12}^T\gamma_{2}^{-1}=a \Psi_{21},
\ee
where $a$ can be $\pm 1$ or $\pm i$.
Thus this just reduces the original degrees of freedom by half.

Let us now consider the fermion spectra in  the D5-D$q$ system.
To find the spectrum, it is important to note that the non-BPS D-branes
do not have any GSO projection and lead to Majorana instead of Weyl fermions.
Also we should note that the D5-D3 system with $\mathrm{O}9^+$-plane (class D)
and the D5-D7 system with $\mathrm{O}9^-$-plane (class C)
should be understood as
the $\Omega$ projection of the $\mathrm{D}5-\mathrm{D}3-\ov{\mathrm{D}3}$
and $\mathrm{D}5-\mathrm{D}7-\ov{\mathrm{D}7}$ system in type IIB string.
Another important fact is that there are two D5-branes coincident
in the case of $\mathrm{O}9^-$-plane
and so we have to double the spectrum in this case.
These considerations lead to
the fermion spectra summarized in Table \ref{AandAIII} and Table \ref{real case}.



\begin{thebibliography}{99}

\bibitem{KaneMele}
C.\ L.\ Kane and E.\ J.\ Mele,
Phys.\ Rev.\ Lett.\ \textbf{95}, 146802 (2005);
\textbf{95}, 226801 (2005).

\bibitem{Roy}
R.\ Roy,
Phys.\ Rev.\ B \textbf{79}, 195321 (2009);
\textit{ibid}, 195322 (2009).

\bibitem{moore07}
J.\ E.\ Moore and L.\ Balents,
Phys.\ Rev.\ B \textbf{75}, 121306(R) (2007).

\bibitem{bernevig06}
B.\ A.\ Bernevig, T.\ L.\ Hughes, and S.-C.~Zhang,
Science \textbf{314}, 1757 (2006).

\bibitem{konig07}
M.\ K\"onig \textit{et al.},
Science \textbf{318}, 766 (2007).

\bibitem{Fu06_3Da}
L.\ Fu, C.\ L.\ Kane, and E.\ J.\ Mele,
Phys.\ Rev.\ Lett.\ \textbf{98}, 106803 (2007).

\bibitem{Fu06_3Db}
L.\ Fu and C.\ L.\ Kane,
Phys.\ Rev.\ B \textbf{76}, 045302 (2007).

\bibitem{Qi2008}
X.-L.\ Qi, T.\ Hughes,  and S.-C.\ Zhang,
Phys.\ Rev.\ B \textbf{78}, 195424 (2008).

\bibitem{hasan07}
D.\ Hsieh \textit{et al},
Nature \textbf{452}, 970 (2008).


\bibitem{Hsieh09}
D.\ Hsieh,
Science \textbf{323}, 919 (2009).


\bibitem{Xia09}
Y. Xia et al,
Nature Phys. \textbf{5}, 398 (2009).

\bibitem{Hsieh09b}
D.\ Hsieh et al,
Nature \textbf{460}, 1101 (2009).


\bibitem{Chen09}
Y.\ L.\ Chen et al,
Science \textbf{325}, 178 (2009).

\bibitem{review}
M. Z. Hasan, and C. L. Kane,
\texttt{arXiv:1002.3895}.



\bibitem{SRFL}
A.\ Schnyder, S.~Ryu, A.~Furusaki, A.\ Ludwig,
Phys.\ Rev.\ B \textbf{78}, 195125 (2008);
AIP Conf. Proc. \textbf{1134}, 10 (2009);
S.\ Ryu, A.\ Schnyder, A.~Furusaki, and A.\ Ludwig,
New J. Phys. \textbf{12} 065010 (2010).



\bibitem{Kitaev}
A.~Kitaev,
AIP Conf.\ Proc. \textbf{1134}, 22 (2009).

\bibitem{HoravaFermiSurfKtheory}
For an application of K-theory to gapless fermion systems with a fermi surface
with the parallel between K-theory classification of D-brane charges,
see
P.\ Horava,
Phys.\ Rev.\ Lett.\ \textbf{95}, 016405 (2005).


\bibitem{Zirnbauer96}
M.\ R.\ Zirnbauer,
J.\ Math.\ Phys.\ \textbf{37}, 4986 (1996).

\bibitem{Altland97}
A.\ Altland and M.\ R.\ Zirnbauer,
Phys.\ Rev.\ B \textbf{55}, 1142 (1997).

\bibitem{Huckleberry2005}
See, e.g., also a more recent discussion:
P.\ Heinzner, A.\ Huckleberry, and M.\ R.\ Zirnbauer,
Commun.\ Math.\ Phys.\ \textbf{257}, 725 (2005).


\bibitem{Helgason1978}
S.\ Helgason,
``Differential geometry, Lie groups and symmetric spaces'',
Academic Press, New York, (1978).

\bibitem{SchnyderRyuLudwig2009}
A.\ P.\ Schnyder,
S.\ Ryu,
and
A.\ W.\ W.\ Ludwig,
Phys.\ Rev.\ Lett. \textbf{102}, 196804 (2009).


\bibitem{interactions}
See for example,
S.\ -S.\ Lee, S.\ Ryu,
Phys.\ Rev.\ Lett.\ \textbf{100}, 186807 (2008);
M.\ W.\ Young, S.\ -S.\ Lee, C.\ Kallin,
Phys.\ Rev.\ B \textbf{78}, 125316 (2008);
L.\ Fidkowski, and A.\ Kitaev,
\texttt{arXiv:0904.2197};
M.\ Levin and A.\ Stern,
Phys.\ Rev.\ Lett. \textbf{103}, 196803 (2009);
D.\ Pesin, and L.\ Balents,
Nature Phys. \textbf{6}, 376 (2010);
S. Rachel, and K. Le Hur,
\texttt{1003.2238};
J.\ Maciejko, X.\ -L.\ Qi, A.\ Karch, S.\ -C.\ Zhang,
\texttt{arXiv:1004.3628};
B.\ Swingle, M.\ Barkeshli, J.\ McGreevy, T. Senthil,
\texttt{arXiv:1005.1076};
J.\ Wen, A.\ Ruegg, C. -C. Joseph Wang, G.\ A.\ Fiete,
\texttt{1005.4061};
C.\ N.\ Varney, K.\ Sun, M.\ Rigol, V.\ Galitski,
\texttt{1007.3502}.


\bibitem{WittenK}
  E.~Witten,
  JHEP {\bf 9812}, 019 (1998).

\bibitem{Hor}
  P.~Horava,
  Adv.\ Theor.\ Math.\ Phys.\  {\bf 2} (1999) 1373.



\bibitem{Sen}
  A.~Sen,
  Int.\ J.\ Mod.\ Phys.\  A {\bf 20}, 5513 (2005).

\bibitem{SenSO}
  A.~Sen,
  JHEP {\bf 9809}, 023 (1998)
  [arXiv:hep-th/9808141].



\bibitem{Karch}
For a different approach to TIs from string theory, see, for example,
  A.~Karch,
  Phys.\ Rev.\ Lett.\  {\bf 103}, 171601 (2009).


\bibitem{Wen91-99}
X.\ -G.\ Wen, Mod.\ Phys.\ Lett.\ B \textbf{5}, 39 (1991);
Phys.\ Rev.\ B \textbf{60}, 8827 (1999).

\bibitem{DbraneShort2010}
S.\ Ryu and T.\ Takayanagi,
\texttt{arXiv:1001.0763} (2010).

\bibitem{Pol}
  J.~Polchinski,
  ``String theory. Vol. 1 and 2'',
Cambridge, UK: Univ. Pr. (1998).




\bibitem{Rey}
  S.~J.~Rey,
  Prog.\ Theor.\ Phys.\  {\bf 177}, 128 (2009).

\bibitem{DKS}
  J.~L.~Davis, P.~Kraus and A.~Shah,
  JHEP {\bf 0811}, 020 (2008);
O.~Bergman, N.~Jokela, G.~Lifschytz and M.~Lippert,
  arXiv:1003.4965 [hep-th].



\bibitem{FujitaFQHE}
M.~Fujita, W.~Li, S.~Ryu, and T.~Takayanagi,
JHEP \textbf{0906}, 066 (2009);
  Y.~Hikida, W.~Li and T.~Takayanagi,
  JHEP {\bf 0907} (2009) 065
 .





\bibitem{ISS}
For a KR theory analysis in similar brane systems, see
T.~Imoto, T.~Sakai and S.~Sugimoto,
\texttt{arXiv:0907.2968},
and references therein.

\bibitem{Pavan09}
Pavan Hosur,
Shinsei Ryu,
and
Ashvin Vishwanath,
Phys.\ Rev.\ B, \textbf{81}, 045120 (2010).



\bibitem{footnote}
For $d=0$, what we mean by Majorana (``Mj'') is
a fermion with some real condition.

\bibitem{Berg}
O.~Bergman,
  ``Tachyon condensation in unstable type I D-brane systems,''
  JHEP {\bf 0011} (2000) 015
  [arXiv:hep-th/0009252].



\bibitem{antibrane}
 A.~Sen,
  JHEP {\bf 9808}, 012 (1998)
  [arXiv:hep-th/9805170].

\bibitem{nonBPS}
O.~Bergman and M.~R.~Gaberdiel,
  ``Stable non-BPS D-particles,''
  Phys.\ Lett.\  B {\bf 441}, 133 (1998)
  [arXiv:hep-th/9806155].


\bibitem{Kitaev05}
Alexei Kitaev,
Ann.\ of Phys.\ \textbf{321}, 2 (2006).

\bibitem{Ryu08}
Shinsei Ryu,
Phys.\ Rev.\ B \textbf{79}, 075124 (2009).

\bibitem{Essin08}
A.\ M.\ Essin, J.\ E.\ Moore and D.\ Vanderbilt,
Phys.\ Rev.\ Lett.\ \textbf{102}, 146805 (2009).



\bibitem{QSHEZ}
B. Andrei Bernevig, Shou-Cheng Zhang,
Phys.\ Rev.\ Lett.\  {\bf 96}, 106802 (2006).


\bibitem{Ryumasses}
Shinsei Ryu, Christopher Mudry, Chang-Yu Hou, and Claudio Chamon
Phys.\ Rev.\ B \textbf{80}, 205319 (2009).

\bibitem{Read00}
N.\ Read and D.\ Green,
Phys.\ Rev.\ B \textbf{61}, 10267 (2000).


\bibitem{SenthilMarstonFisher}
T.\ Senthil, J.\ B. Marston, and M.\ P.\ A. Fisher,
Phys.\ Rev.\ B \textbf{60}, 4245 (1999);
I.\ A.\ Gruzberg,
A.\ W .\ W.\ Ludwig,
and N.\ Read
Phys.\ Rev.\ Lett.\ \textbf{82}, 4524 (1999).

\bibitem{HoyosBadajozJensenKarch2010}
Carlos Hoyos-Badajoz, Kristan Jensen, Andreas Karch,
\texttt{arXiv:1005.3253}.

\bibitem{RTEE}
 S.~Ryu and T.~Takayanagi,
  Phys.\ Rev.\ Lett.\  {\bf 96} (2006) 181602;
  JHEP {\bf 0608} (2006) 045;
 T.~Nishioka, S.~Ryu and T.~Takayanagi,
  J.\ Phys.\ A  {\bf 42} (2009) 504008.

\bibitem{TopEE}
A.~Kitaev and J.~Preskill,
Phys.\ Rev.\ Lett.\ \textbf{96}, 110404 (2006);
M.~Levin and X-G.~Wen,
Phys.\ Rev.\ Lett.\ \textbf{96}, 110405 (2006).



















\end{thebibliography}
\end{document}